\documentclass[
 reprint,
 amsmath,amssymb,
 aps,
 prl,
]{revtex4-2}

\usepackage{graphicx}
\usepackage{dcolumn}
\usepackage{bm}
\usepackage{braket}
\usepackage{mathtools}
\usepackage{upgreek}

\usepackage[
    colorlinks=true,
    citecolor=blue,
    linkcolor=blue,
    urlcolor=blue
]{hyperref}

\begin{document}

\preprint{APS/123-QED}

\title{Dipolar Exciton Polarons in a van der Waals Homobilayer}

\author{Yuxin Duan}
\thanks{These authors contributed equally to this work.}
\author{Hongming Zhang}
\thanks{These authors contributed equally to this work.}
\author{Yaofeng Zhu}
\author{Hang Zhou}
\author{Xinyue Zhang}
\author{Yan Liu}
\author{Song Luo}
\author{Long Zhang}
\email{zhanglong@xmu.edu.cn}
\author{Zhanghai Chen}
\email{zhanghai@xmu.edu.cn}

\affiliation{Department of Physics, College of Physical Science and Technology,Xiamen University, Xiamen 361005, China
}

\date{\today}

\begin{abstract}
Fermi polarons exemplify how many-body interactions give rise to emergent quasiparticles, providing a fundamental framework for understanding interacting quantum matter. In electron-doped two-dimensional semiconductors, exciton polarons offer optical access to Fermi-polaron physics and electrical tunability, yet achieving stronger and more controllable exciton–electron interactions remains a key challenge. Here, we realize dipolar exciton polarons in an electrically tunable van der Waals homobilayer, where interlayer excitons with spatially separated electrons and holes act as quantum impurities carrying permanent electric dipole moments. These dipolar polarons exhibit substantially enhanced exciton–electron interactions, with binding energies more than four times those of conventional intralayer exciton polarons. We further demonstrate that Fermi-sea layer polarization provides an additional degree of freedom for controlling many-body interactions. In particular, the presence of a Fermi sea induces hybridization between oppositely oriented dipolar polarons, whereas confining the Fermi sea to a single layer suppresses this hybridization and produces orientation-selective many-body dressing. Our findings establish an experimental platform for dipolar-polaron physics in which dipole orientation and Fermi-sea layer polarization jointly govern many-body interactions, opening a route towards electrically tunable interactions and dipolar-polaron-mediated attraction between itinerant electrons.

\end{abstract}

\maketitle


\section{\label{sec:level1} Introduction}

Fermi polarons, quasiparticles formed when mobile impurities are dressed by particle–hole excitations of a surrounding Fermi sea, provide a paradigmatic setting for understanding how well-defined quasiparticles emerge in strongly interacting quantum many-body systems \cite{Combescot2007,Schirotzek2009,Massignan2014}. Beyond the single-impurity problem, their attractive and repulsive branches are closely connected to pairing \cite{Punk2009,Koschorreck2012} and magnetic instabilities \cite{Pilati2010,Massignan2014}, providing microscopic insights into collective phenomena in interacting Fermi systems \cite{ChevyMora2010}. Fermi polarons therefore serve as versatile probes of strongly correlated quantum matter across diverse physical platforms.

In recent years, atomically thin transition-metal dichalcogenides (TMDs) have emerged as a powerful platform for Fermi-polaron physics \cite{Sidler2017,Efimkin2017,Liu2021,Smolenski2019,Cui2024InterlayerPolaron}. In these materials, tightly bound excitons act as optically accessible quantum impurities, while electrostatic gating enables continuous control of the surrounding electron or hole Fermi sea. This combination has enabled the observation of attractive and repulsive exciton polarons \cite{Sidler2017} and the exploration of their quantum dynamics \cite{Ni2025,Huang2023}, valley-dependent many-body dressing \cite{Cong2023}, and hybridization with cavity photons \cite{Tan2020,Chervy2020,BastarracheaMagnani2021,Zhang2026Landau}. Resonant exciton–electron interactions further provide access to strongly tunable many-body scattering \cite{Kuhlenkamp2022,Wagner2025}. Despite this progress, experimentally realized exciton polarons have predominantly involved impurities whose interactions with the surrounding Fermi sea are effectively short-ranged or spatially isotropic \cite{Massignan2026}. Extending exciton-polaron physics into a regime governed by permanent electric dipoles would introduce intrinsically long-range and anisotropic interactions, with their character determined by the relative spatial orientation of the interacting particles \cite{Hubert2019,SanchezBaena2024}.

Dipolar interactions have attracted substantial theoretical interest because of their potential to support anisotropic quasiparticle dressing and unconventional interaction channels, including both attractive and repulsive interactions \cite{Lahaye2009}. Dipolar polarons may consequently provide a versatile platform for exploring complex many-body states, including bipolaron formation \cite{CamachoGuardian2018}, which has been proposed as a possible precursor to superconductivity \cite{Zhang2023Bipolaron}. Despite extensive theoretical proposals and significant advances in dipolar quantum systems \cite{Valtolina2020,Schindewolf2022,Bigagli2024}, Fermi polarons formed by impurities carrying permanent electric dipole moments remain largely unexplored experimentally \cite{Massignan2026}. Realizing such quasiparticles would therefore extend the accessible landscape of Fermi-polaron physics and enable the investigation of how dipolar interactions and the structure of the surrounding Fermi sea jointly determine many-body behavior~\cite{Qi2025Trion,Nguyen2025Trion}.

Here, we realize dipolar exciton polarons in an electrically tunable van der Waals homobilayer. Interlayer excitons, in which electrons and holes reside in different layers, act as quantum impurities carrying permanent out-of-plane electric dipole moments. These dipolar polarons exhibit substantially enhanced exciton–electron interactions, with binding energies more than four times those of conventional intralayer exciton polarons. Beyond establishing this new impurity regime, we demonstrate that the layer distribution of the Fermi sea provides an additional degree of freedom for controlling many-body interactions. Specifically, we demonstrate Fermi-sea-assisted hybridization between oppositely oriented dipolar polarons. This hybridization involves interlayer hole tunnelling and electron exchange between the excitons and the Fermi sea. Confining the Fermi sea to a single layer suppresses this hybridization and gives rise to orientation-selective many-body dressing. Our results establish a solid-state platform for exploring long-range and anisotropic polaron physics, in which dipole orientation and Fermi-sea layer polarization jointly govern many-body interactions. This tunable platform opens opportunities to investigate dipolar-polaron-mediated attraction between itinerant electrons and the emergence of correlated phases in interacting Bose–Fermi systems. Coupling these dipolar polarons to cavity photons could further enable tunable photon-photon interactions and nonclassical light generation\cite{Xiang2026,Kang2026}.

\section{}

\begin{figure}[htbp]    
    \centering          
    \includegraphics[width=\columnwidth]{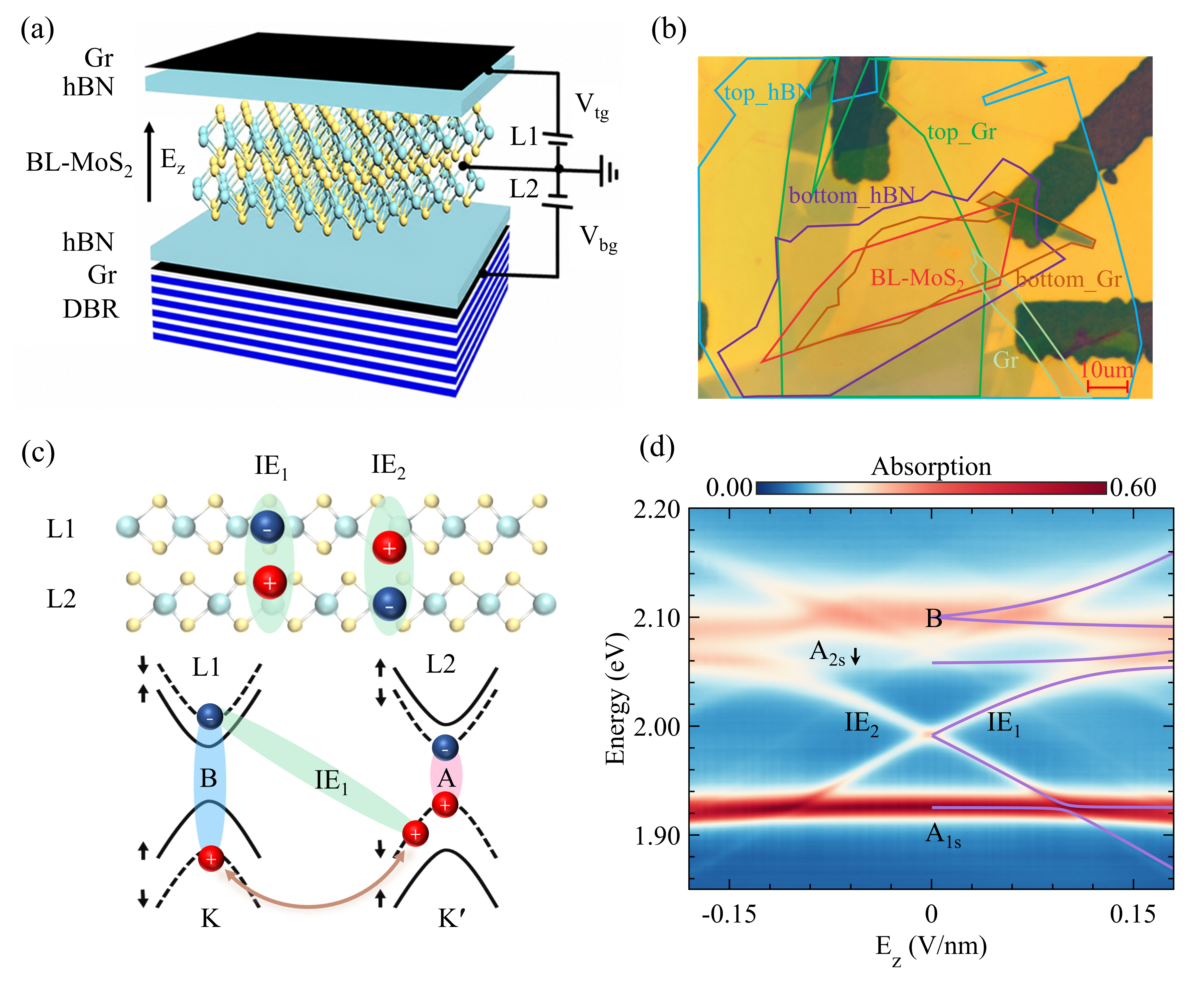}  
    \caption{Electrically tunable interlayer excitons in bilayer MoS\textsubscript{2}. (a) Schematic of the dual-gate MoS\textsubscript{2} bilayer device on a DBR substrate. (b) Optical micrograph of the device. Colored contours indicate the constituent flakes, as labeled. Scale bar $10~\upmu\mathrm{m}$.(c) Top panel: Schematic real-space configurations of IE\textsubscript{1} and IE\textsubscript{2}, corresponding to opposite electron-hole layer configurations. Blue (red) spheres denote electrons (holes), and L1, L2 label the two MoS\textsubscript{2} layers. Bottom panel: Band diagram illustrating the intralayer A and B excitons and the interlayer exciton (IE). Solid and dashed curves denote spin-up and spin-down bands, respectively; the brown curved arrow indicates interlayer hole tunneling.(d) Electric field dependent absorption spectra of the bilayer MoS\textsubscript{2} device, with the purple solid lines showing fits from a multiple coupled-oscillator model. }   
    \label{fig:1}       
\end{figure}

The 2H bilayer MoS\textsubscript{2} device is incorporated into a dual-gate geometry, with hBN layers serving as dielectric spacers and few-layer graphene as the top and bottom gate electrodes , as shown in Fig.~\hyperref[fig:1]{1(a)}. The two gate voltages independently control the out-of-plane electric field $E_z$ and the total carrier density $n$, allowing the electric-field and density dependence of the excitonic states to be investigated separately. The distributed Bragg reflector (DBR) substrate suppresses transmission over the spectral range of interest, allowing the absorption to be determined from the measured reflectance as $A = 1-R$. More details on the device fabricartion and the calibration of electric field and carrier density are presented in Sec.~I of the Supplemental Material~\cite{SupplementalMaterial}.  Figure~\hyperref[fig:1]{1(b)} shows the microscope image of the device.  Figure~\hyperref[fig:1]{1(c)} shows the relevant excitonic configurations in 2H bilayer MoS\textsubscript{2}. We denote the interlayer exciton with the electron in the top layer and the hole predominantly in the bottom layer as IE\textsubscript{1}, while the opposite configuration is labeled IE\textsubscript{2}. The two states therefore carry permanent out-of-plane electric dipole moments with opposite orientations. The opposite dipole orientations of IE\textsubscript{1} and IE\textsubscript{2} are directly revealed by their electric-field dependence. As shown in the absoription spectra in Fig.~\hyperref[fig:1]{1(d)}, the two interlayer exciton branches are  degenerate around $E_z = 0$ and shift approximately linearly in opposite directions with increasing electric field, the electric dipole moments of IE\textsubscript{1} and IE\textsubscript{2} are extracted from their linear Stark effect, as detailed in Sec.~II of the Supplemental Material~\cite{SupplementalMaterial}. In contrast,  the intralayer B exciton, and both A\textsubscript{1s} exciton and its Rydberg series A\textsubscript{2s}, show negligible field dependence. 

Owing to strong interlayer hybridization of the valence bands, the hole component is partially delocalized across the two layers \cite{Gerber2019,Sponfeldner2022}. This hybridization allows the interlayer excitons to couple with optically bright intralayer B excitons, giving them appreciable oscillator strength \cite{Leisgang2020,Peimyoo2021,Shree2021,Lorchat2021,Louca2023,Horng2018}.  Taking hybridization between different type of interlayer and intralayer excitons, the electric field dependence of absorption spectra are well reproduced by a multiple coupled-oscillator model, as shown in Fig.~\hyperref[fig:1]{1(d)} (see Secs.~III and IV of the Supplemental Material~\cite{SupplementalMaterial} for details of the model and the assignment of additional interlayer exciton species). These measurements establish the MoS\textsubscript{2} homobilayer as a system hosting electrically tunable interlayer excitons with opposite out-of-plane dipole configurations, providing a controllable dipolar quantum impurity setting for studying many-body dressing by an electron Fermi sea.
 

\begin{figure}[htbp]    
    \centering          
    \includegraphics[width=\columnwidth]{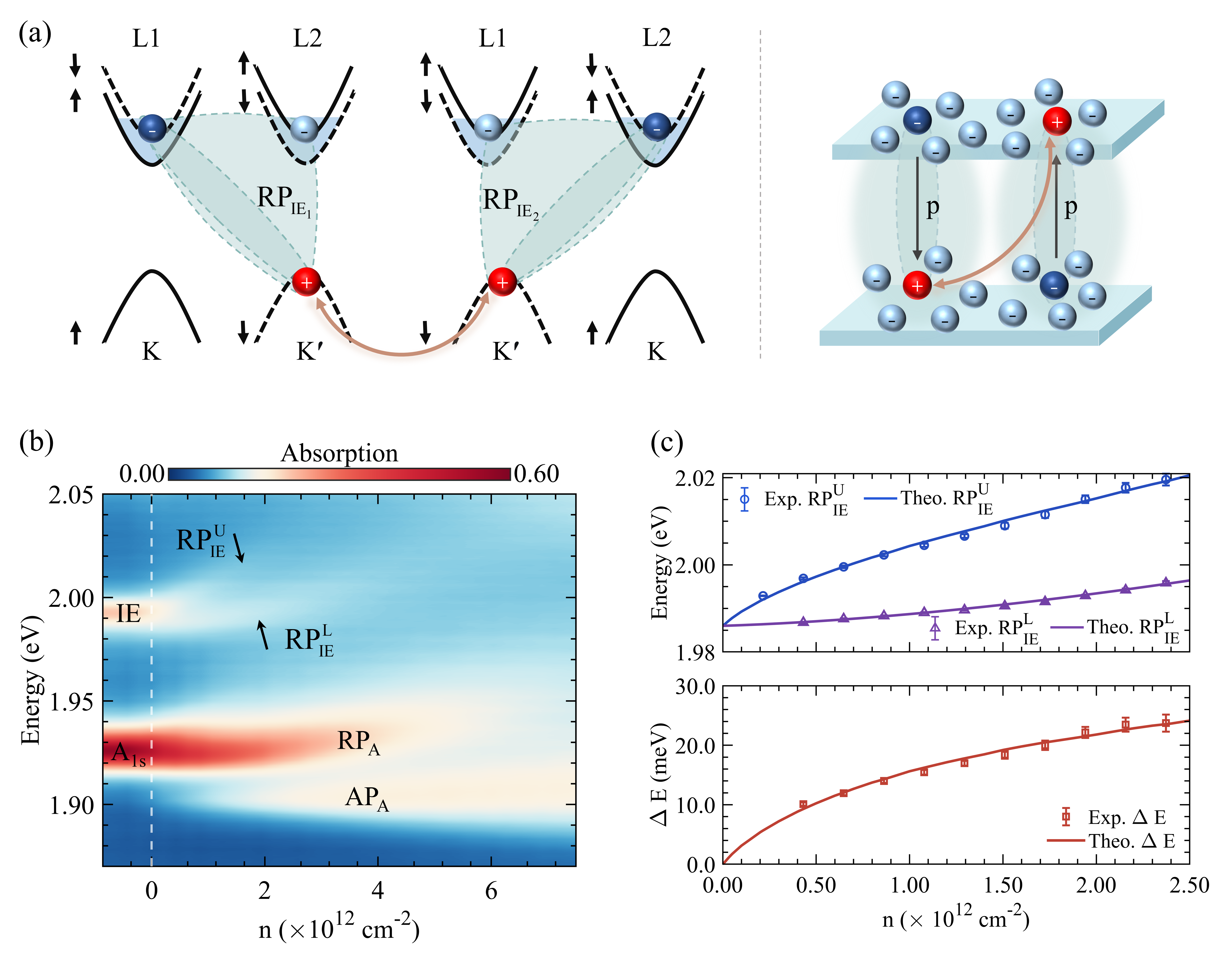}  
    \caption{Fermi-sea-induced hybridization of oppositely oriented dipolar polarons. (a)Schematic illustration of the Fermi-sea-mediated coupling between two interlayer polaron configurations with opposite dipole orientations. Left panel: Band diagram showing the formation of two interlayer repulsive polaron configurations through the dressing of interlayer excitons by the Fermi seas in their respective hole layers. Interlayer hole tunneling couples the two polaron configurations. Right panel: Real-space illustration of interlayer excitons with opposite out-of-plane dipole moments interacting with electron Fermi seas in the two layers.(b) Absorption spectra of bilayer MoS\textsubscript{2} as a function of electron density n. (c) Extracted energies of $\mathrm{RP}_{\mathrm{IE}}^{\mathrm{H}}$ and $\mathrm{RP}_{\mathrm{IE}}^{\mathrm{L}}$ as a function of electron density n. The experimental results (symbols) are compared with theoretical calculations (solid curves). The energy separation $\Delta E = E_{\mathrm{RP}_{\mathrm{IE}}^{\mathrm{H}}} - E_{\mathrm{RP}_{\mathrm{IE}}^{\mathrm{L}}}$. }   
    \label{fig:2}       
\end{figure}

To investigate how dipolar excitons interact with an electron Fermi sea, we electrostatically dope the bilayer MoS\textsubscript{2} while maintaining $E_z \approx 0$. The corresponding electron density dependent absorption spectra are shown in Fig.~\hyperref[fig:2]{2(b)}. The intralayer A\textsubscript{1s} exciton evolves into attractive- and repulsive-polaron branches (AP\textsubscript{A} and RP\textsubscript{A}), accompanied by a continuous blueshift of the RP\textsubscript{A} and spectral weight transfer from the RP\textsubscript{A} to the AP\textsubscript{A} with increasing electron density\cite{Sidler2017,Efimkin2021,Roch2019,Klein2021}. This  characteristic response provides a clear spectroscopic signature of the tunable electron Fermi sea in our device.

\begin{figure}[htbp]    
    \centering          
    \includegraphics[width=\columnwidth]{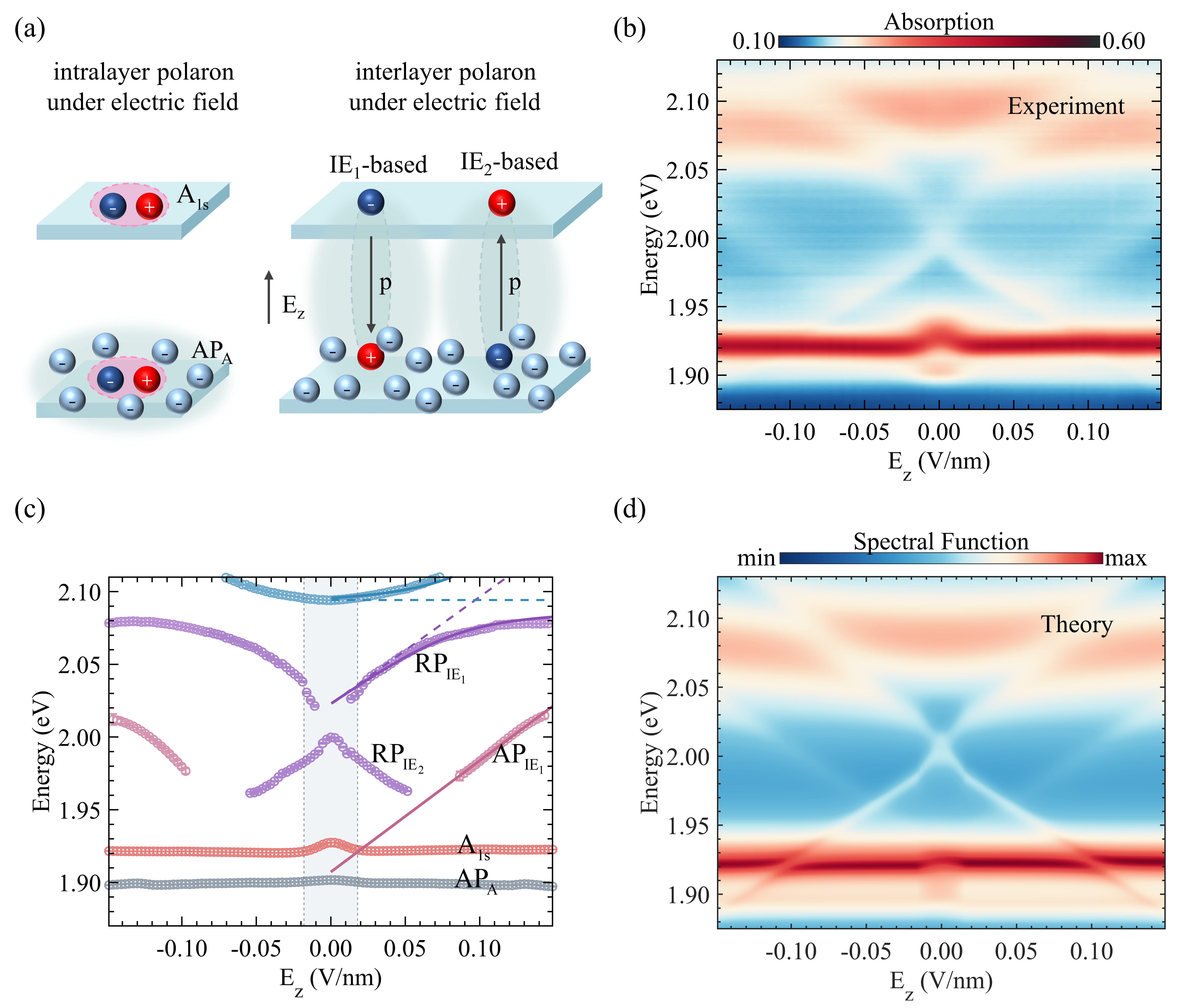}  
    \caption{Orientation-selective polaron dressing by a layer-polarized Fermi sea. (a) Schematic illustration of electric-field-induced layer polarization of the electron Fermi sea and its influence on exciton polarons. Left panel: Evolution of intralayer exciton polarons as electrons redistribute between the two layers. Right panel: Interlayer exciton polarons with opposite dipole orientations interacting with a layer-polarized Fermi sea. (b) Absorption spectra of electron doped bilayer MoS\textsubscript{2} as a function of the out-of-plane electric field E\textsubscript{z} at a fixed electron density  $n \approx 3.2\ \mathrm{\times 10^{12} cm^{-2}}$. (c) Extracted resonance energies of the intralayer and interlayer exciton-polaron branches. The blue and purple solid curves represent fits based on a coupled-oscillator model describing the hybridization between the repulsive interlayer polaron and the intralayer B exciton, while the dashed curves indicate the corresponding uncoupled energy levels. The red solid line represents a linear fit to the Stark shift of the attractive interlayer polaron. The vertical dashed line marks the characteristic electric field corresponding to complete Fermi-sea layer polarization. (d) Theoretically calculated absorption spectra corresponding to the experimental conditions in (b). }   
    \label{fig:3}       
\end{figure}

In contrast, the interlayer exciton exhibits a strikingly different density dependence. With increasing doping density, two well-resolved branches emerge, both shifting toward higher energies and gradually losing spectral weight with increasing electron density, as shown in Fig.~\hyperref[fig:2]{2(b)}. We identify them as repulsive interlayer polaron branches originating from the dressing of the two oppositely oriented interlayer excitons by the electron Fermi sea, and denote the upper- and lower-energy branches as $\mathrm{RP}_{\mathrm{IE}}^{\mathrm{U}}$ and $\mathrm{RP}_{\mathrm{IE}}^{\mathrm{L}}$, respectively. The corresponding attractive polaron branches are not clearly resolved at $E_z\approx 0$ due to spectral overlap with neighboring intralayer excitonic resonances, but can be spectrally resolved through electric-field-induced Stark shifts, as discussed in Fig.~\ref{fig:3}.

To quantify this evolution, we extract the resonance energies of $\mathrm{RP}_{\mathrm{IE}}^{\mathrm{U}}$ and $\mathrm{RP}_{\mathrm{IE}}^{\mathrm{L}}$ and their energy separation $\Delta E$. As shown in Fig.~\hyperref[fig:2]{2(c)}, both branches blueshift with increasing electron density, whereas $\mathrm{RP}_{\mathrm{IE}}^{\mathrm{H}}$ exhibits a substantially larger energy shift, leading to a monotonic increase of $\Delta E$. Under the nearly layer-symmetric condition at $E_z \approx 0$, such a density-dependent splitting can not be understood by independent self-energy renormalization of the two interlayer polarons, indicating an additional coupling between the two quasi-particles.

The origin of this coupling follows from the layer structure of the interlayer excitons [Fig.~\hyperref[fig:2]{2(a)}]. IE\textsubscript{1} and IE\textsubscript{2} correspond to opposite electron-hole layer configurations. Under charge-neutral conditions, spin and momentum conservation constraints suppress the coupling between oppositely oriented interlayer excitons, preventing the emergence of a resolvable anticrossing\cite{Leisgang2020,Peimyoo2021,Liu2025InterlayerCoherence}. When electrons populate both layers, interlayer hole tunneling, together with electron exchange between the exciton and the Fermi sea, couples the two opposite dipole configurations. This combined process conserves total momentum and angular momentum. The resulting hybridization gives rise to the two repulsive-polaron branches, $\mathrm{RP}_{\mathrm{IE}}^{\mathrm{U}}$ and $\mathrm{RP}_{\mathrm{IE}}^{\mathrm{L}}$.

To account for this behavior quantitatively, we calculate the Green function of interlayer exciton dressed by double layer Fermi sea (see Sec.~VII of the Supplemental Material~\cite{SupplementalMaterial}
for details of the theoretical model and calculations):
\begin{equation}
\begin{split}
G^R_{IE}(E) = -\frac{1}{\pi}\,
\mathrm{Im}\Bigg\{
\Bigg[
&\begin{pmatrix}
E-E_{IE_l}-i\gamma & 0 \\
0 & E-E_{IE_{l''}}-i\gamma
\end{pmatrix}
\\
&-
\Sigma^R(E)
\Bigg]^{-1}
\Bigg\}.
\end{split}
\label{eq:spectral_function}
\end{equation}
Where $E_{IE_l}$ is the energy of interlayer exciton with hole locating  in layer $l_l$ or $l_2$ .$\gamma$ is the linewidth of interlayer exciton. $\Sigma^R$ is the self energy matrix of interlayer exciton. To obtain the self energy, we first calculate the scattering matrix under ladder approximation:
\begin{equation}
\begin{split}
T^{R}_{ll',q}(p,p';E)
={}&
U_{ll',q}(p,p')
\\
&+
\sum_{l''}
\int
\frac{d^{2}k}{2\pi^{2}}\,
U_{ll',q}(p,k)
\frac{1-f_{l''}(k)}
{D_{l'',q}(k;E)}
\\
&\qquad\times
T^{R}_{l''l',q}(k,p';E).
\end{split}
\label{eq:Tmatrix}
\end{equation}
where $U_{ll',q}(p,p')=\begin{pmatrix} V_l & W \\ W & V_{l'}\end{pmatrix}$ is the interaction kernel, V is the interaction energy induced by long range dipole-electron. p and k denote the momentum of electrons. f(k) is Fermi distribution function. $\displaystyle\frac{1}{D_{l,p}}=\frac{1}{E-E_{IE_{l}}+\epsilon_e(q)-\epsilon_e(p)-\epsilon_{IE}(q-p)+i\eta}$ is the intermediate state propagator, where $\epsilon_i$ labels kinetic energy of electron or exciton. Then self energy matrix is ontained by integrating the scatterimg matrix, with its explicit expression provided in Sec. VII of the Supplemental Material.
The theoretical results (solid curves in Fig.~\hyperref[fig:2]{2(c)}) reproduce the density evolution of both repulsive polaron branches and their increasing energy separation $\Delta E$. This agreement supports the coupling between oppositely oriented dipolar polarons, enabled by the combined effects of Fermi-sea dressing and interlayer hole tunnelling.

To further investigate the electrical tunability of dipolar exciton polarons, we apply an out-of-plane electric field to control both the interlayer exciton energies and the layer distribution of the electron Fermi sea, as illustrated in Fig.~\hyperref[fig:3]{3(a)}. The electric field lifts the degeneracy of the two oppositely oriented interlayer excitons through the Stark effect, while simultaneously polarizing electrons to a single layer and thereby modifying their many-body interactions. Figure~\hyperref[fig:3]{3(b)} shows the electric-field-dependent absorption spectra at a fixed electron density of $n \approx 3.2 \times 10^{12}\ \mathrm{cm^{-2}}$. The extracted resonance energies are summarized in Fig.~\hyperref[fig:3]{3(c)}, revealing the evolution of both intralayer and interlayer exciton polarons. Additional electric-field-dependent absorption spectra measured at other fixed electron densities are presented in Sec.~VI of the Supplemental Material~\cite{SupplementalMaterial}.

We first use the intralayer exciton resonances as an independent probe of the Fermi-sea layer polarization. As the electric field increases, the repulsive intralayer exciton-polaron branch gradually evolves into the bare A exciton, indicating electron depletion in one layer, whereas the lower-energy resonance retains its attractive-polaron character. These two resonances can therefore be assigned to the bare A exciton in the electron-depleted layer and the attractive exciton polaron in the electron-rich layer, respectively. The latter assignment is consistent with the attractive-polaron branch observed at high electron densities in Fig.~\hyperref[fig:2]{2(a)}. This contrasting behavior provides a spectroscopic signature of complete Fermi-sea layer polarization, with the corresponding electric field marked by the vertical dashed line in Fig.~\hyperref[fig:3]{3(c)}.

We now turn to the dipolar exciton polarons. As the electric field increases, the two oppositely oriented dipolar polaron configurations become increasingly detuned, progressively suppressing their hybridization. More importantly, the attractive branch associated with the configuration whose hole resides predominantly in the electron-rich layer becomes spectrally resolved. Under the electric field, this branch shifts to higher energies due to the Stark effect, becoming spectrally separated from the intralayer A exciton and allowing us to clearly identify the attractive branch. Fitting its electric-field dependence (red solid line) yields an effective electric dipole moment of $(0.77 \pm 0.01)~\mathrm{e}\cdot\mathrm{nm}$. We further extract the effective dipole moment of the corresponding repulsive branch using a coupled-oscillator model that accounts for its hybridization with the intralayer B exciton (dashed line in Fig.~\hyperref[fig:3]{3(c)}), yielding a value of $(0.75 \pm 0.03)~\mathrm{e}\cdot\mathrm{nm}$. The agreement between the two dipole moments further supports the assignment of the attractive and repulsive branches to the same interlayer exciton configuration. The energy separation between the attractive and repulsive dipolar polaron branches, extrapolated to zero electric field, yields a polaron binding energy of $(115.5 \pm 1.2)~\mathrm{meV}$, more than four times that of intralayer exciton polarons ($(25.7 \pm 0.1)~\mathrm{meV}$) under the same conditions. This substantially enhanced binding indicates stronger many-body interactions between dipolar excitons and the Fermi sea, which can be phenomenologically understood from the spatial separation of the electron and hole, with the hole residing in the electron-rich layer. Our many-body calculations quantitatively reproduce this enhanced binding energy, as shown in Fig.~\hyperref[fig:3]{3(d)}.

\begin{figure}[htbp]    
    \centering          
    \includegraphics[width=\columnwidth]{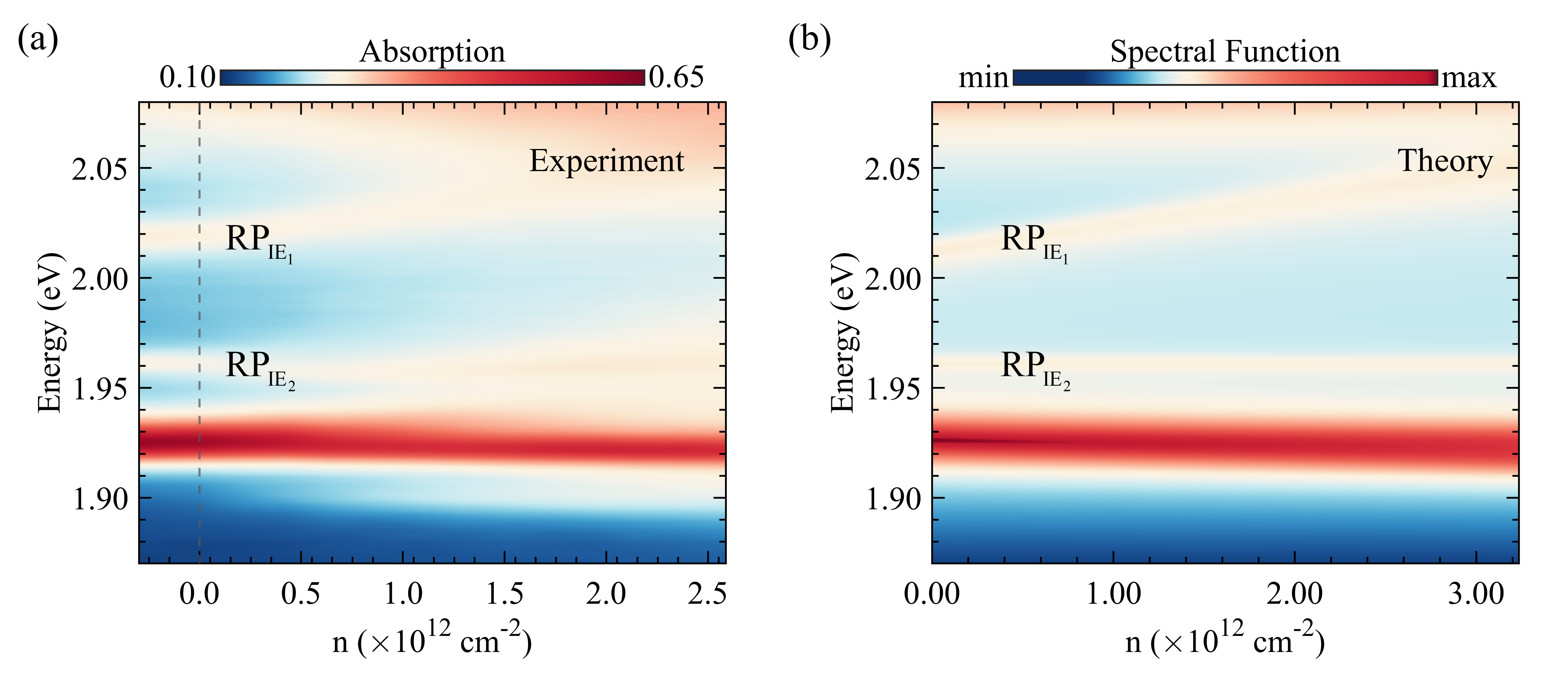}  
    \caption{Electron-density dependence of oppositely oriented dipolar polarons. (a) Reflectance-contrast spectra of electron doped bilayer MoS\textsubscript{2} measured at a fixed out-of-plane electric field $E_{z} \approx 0.049\ \mathrm{V/nm} $. (b) Calculated absorption spectra under the same conditions as in (a).
}   
    \label{fig:4}       
\end{figure}

To independently investigate the many-body response of the two oppositely oriented dipolar polarons to the electron Fermi sea, we measure their density dependence at a fixed out-of-plane electric field. We choose $E_z \approx 0.049\ \mathrm{V/nm}$, where the two dipolar polaron configurations are effectively decoupled and the electron Fermi sea remains polarized into the bottom layer throughout the investigated density range. Figure~\hyperref[fig:4]{4(a)} shows the corresponding absorption spectra of bilayer MoS$_2$. At charge neutrality, the degeneracy between IE$_1$ and IE$_2$ is lifted by the Stark effect. Upon electron doping, the two interlayer polaron branches exhibit strikingly different density dependences. $\mathrm{RP}_{\mathrm{IE}_1}$ continuously blueshifts with increasing $n$, whereas $\mathrm{RP}_{\mathrm{IE}_2}$ remains nearly unchanged over a broad density range. This asymmetric response arises from the different layer configurations of the dipolar polarons relative to the Fermi sea. $\mathrm{RP}_{\mathrm{IE}_1}$ originates from the interlayer exciton configuration whose hole resides predominantly in the electron-rich layer and therefore acquires a strong density-dependent many-body self-energy. In contrast, $\mathrm{RP}_{\mathrm{IE}_2}$ derives from the opposite dipolar configuration, whose hole resides predominantly in the electron-depleted layer, and is only weakly renormalized. The calculated spectral function in Fig.~\hyperref[fig:4]{4(b)} reproduces the contrasting density dependence of the two polaron branches, further supporting the layer-dependent many-body dressing picture. These results demonstrate that the many-body response of dipolar exciton polarons depends strongly on their orientation relative to the layer-polarized Fermi sea.

We have realized dipolar exciton polarons in bilayer MoS\textsubscript{2} and demonstrated electrical control of their hybridization and spectral response through the layer polarization of the Fermi sea. These findings establish the relative layer configuration of the exciton and its electronic environment as a control parameter for polaron physics. These results establish bilayer MoS\textsubscript{2} as an electrically reconfigurable quantum impurity platform in which dipole configuration and Fermi sea layer polarization jointly control many-body dressing, opening new opportunities for exploring correlated quantum states governed by coupled charge, layer, and dipolar degrees of freedom in van der Waals materials\cite{Zhang2022ExcitonInsulator,Zhang2023Dipolar,Yu2023Quadrupolar}.Time-resolved coherent spectroscopy could reveal how these quasiparticles form and whether coherent evolution between opposite dipole configurations survives coupling to the electronic environment. Tuning the Fermi-sea layer polarization could then allow the competition between carrier-mediated coupling and decoherence to be explored, providing access to the nonequilibrium dynamics of dipolar quantum impurities.


\begin{acknowledgments}

\end{acknowledgments}


\bibliography{apssamp}
\bibliographystyle{apsrev4-2}
\end{document}